# Mid-infrared dual-comb spectroscopy with electro-optic modulators

Ming Yan[1,2], Pei-Ling Luo[1,2], Kana Iwakuni[1,2], Guy Millot[3], Theodor W. Hänsch[1,2] and Nathalie Picqué[1,2,4,*]

1. Max-Planck-Institut für Quantenoptik, Hans-Kopfermann-Straße 1, 85748 Garching, Germany
2. Ludwig-Maximilians-Universität München, Fakultät für Physik, Schellingstr. 4/III, 80799 München, Germany
3. Laboratoire Interdisciplinaire Carnot de Bourgogne, UMR 6303 CNRS - Univ. Bourgogne Franche-Comté, 9 Avenue A. Savary, F-21078 Dijon, France
4. Institut des Sciences Moléculaires d'Orsay (ISMO), CNRS, Univ. Paris-Sud, Université Paris-Saclay, F-91405 Orsay, France
*Corresponding author: nathalie.picque@mpq.mpg.de

**Abstract:** We demonstrate dual-comb spectroscopy based on difference frequency generation of frequency-agile near-infrared frequency combs, produced with the help of electro-optic modulators. The combs have a remarkably flat intensity distribution and their positions and line spacings can be selected freely by simply dialing a knob. We record, in the 3-µm region, Doppler-limited absorption spectra with resolved comb lines within milliseconds. Precise molecular line parameters are retrieved. Our technique holds promise for fast and sensitive time-resolved studies e.g. of trace gases.

Laser frequency combs [1], and their broad spectrum of evenly spaced narrow lines with precisely known positions, open up novel opportunities for broadband molecular spectroscopy. In the last decade, a number of promising new techniques and experimental schemes interrogating the sample with a frequency comb have been demonstrated and perfected [2-9]. Amongst such techniques, dual-comb spectroscopy -which measures the time-domain interference between two combs of slightly different line spacings- has the distinguishing advantage of a multiplex instrument without moving parts. Similar to a Michelson-based Fourier transform spectrometer, all the spectral elements are simultaneously measured on a single photo-detector. Unlike the Michelson interferometer, however, the phase difference is automatically scanned with a static device. This potentially enables fast measurement speed and high resolution in virtually any spectral region from the THz range to the extreme ultraviolet, as well as self-calibration of the frequency scale, up to the accuracy of an atomic clock. Dual-comb spectroscopy also presents specific instrumental challenges that need to be overcome before the technique can reach its full potential. The mid-infrared spectral range, where most molecules in the gas phase have strong fundamental ro-vibrational transitions, is extremely useful for precise high-resolution laboratory spectroscopy and for applications to e.g. trace gas detection. Laser frequency comb generation here is still a technology in active development [10-15]. Many promising proof-of-principle experiments of dual-comb spectroscopy have been demonstrated with a variety of frequency comb sources including quantum cascade lasers directly emitting in the mid-infrared [16,17], solid state lasers emitting at the edge of the mid-infrared region [18], synchronously pumped optical parametric oscillators [19,20] and laser systems based on difference frequency generation [2,3,21,22]. However, spectra at Doppler-limited resolution have only been reported once [21], around 3 µm, with fully optically-stabilized laser systems. The determination of precise molecular line parameters has remained limited to accurate line position measurements [21] in the $\nu_3$ band of methane. In this article, we report a new approach to real-time mid-infrared Doppler-limited dual-comb spectroscopy. We implement such approach in the region between 85 and 95 THz (between 3150 and 3500 nm), an atmospheric window important for the detection of hydrocarbons as well as oxygen- or nitrogen-containing organic compounds. Precise line parameters, including positions and intensities, are retrieved from spectra measured on the millisecond time scale.

We describe a set-up dedicated to mid-infrared dual-comb spectroscopy. Laser frequency comb oscillators are not readily available in the 85-95-THz region and dual-comb systems based on nonlinear frequency conversion of mode-locked lasers may be complex and of limited agility. Therefore, we take a different approach, which involves difference frequency generation of frequency combs based on electro-optic modulators. The experimental setup is illustrated in Fig. 1a. We first generate two near-infrared frequency-agile frequency combs in the telecommunication region with a fibered set-up, as already described in [23]. The output of a continuous-wave laser diode (CTL 1550, Toptica), with an optical frequency $f_{signal}$ tunable from 185 to 196 THz (1530 to 1620 nm) at a tuning speed of 1.2 THz s$^{-1}$ (10 nm s$^{-1}$), is split into two arms. In one arm, we shift the frequency of the continuous-wave laser to $f_{signal}+f_{shift}$ with an acousto-optic frequency shifter driven at the radio-frequency $f_{shift}$=25MHz. Then, the intensity of each beam is modulated by an electro-optic modulator at a repetition frequency of $f$ and $f+\Delta f$, respectively. The repetition frequencies $f$ and $f+\Delta f$ are chosen between 100 MHz and 500 MHz and they are set by radio-frequency synthesizers. A straightforward change of electronic filters would shift this frequency range to lower or higher repetition frequencies. After amplification in erbium-doped fiber amplifiers, two asynchronous trains of pulses with an average power of 200 mW and a pulse duration of 50 ps are produced. For spectral broadening, the two pulse trains counter-propagate in a 1.3-km-long dispersion-compensating nonlinear fiber. The fiber has a high and flat normal dispersion ($D$=-94 ps nm$^{-1}$ km$^{-1}$ at 191.1 THz – 1569 nm) and a nonlinear coefficient of 3 W$^{-1}$km$^{-1}$. When the line spacing $f$ is 300 MHz, the two near-infrared frequency combs comprise 2N+1=1201 lines each and span 360 GHz (Fig. 1b). For a constant average power, the number of comb lines increases when the repetition frequency decreases. The frequency of each comb line may be written $f_{signal}+nf$ and $f_{signal}+f_{shift}+nf+\Delta f$, respectively, where n is an integer varying between $-N$ and $+N$. More details about the near-infrared frequency comb generation may be found in [23].

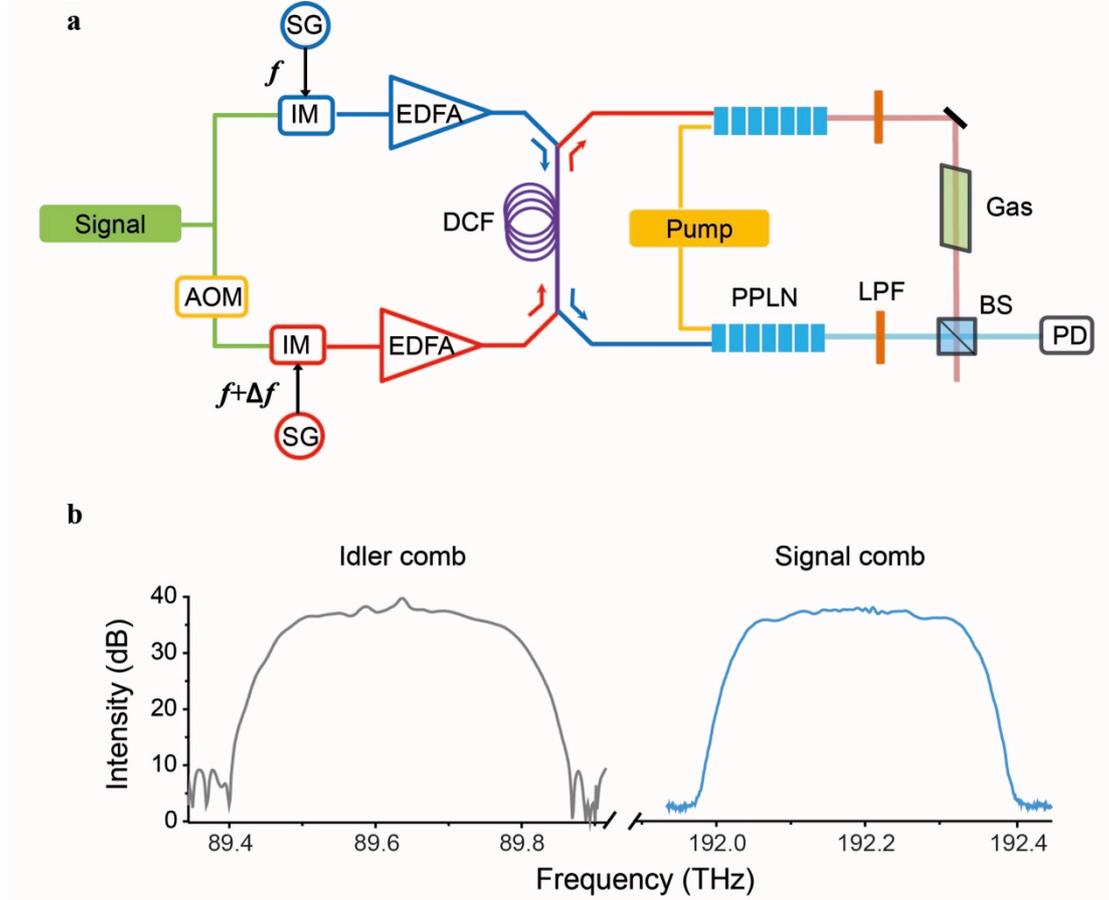

**Figure 1. Mid-infrared flat-top frequency comb generation.**
(a) Experimental setup. AOM: acousto-optic frequency shifter; SG: radio-frequency signal generator; IM: intensity modulator; EDFA: erbium-doped fiber amplifier; DCF: dispersion compensating fiber; PPLN: Magnesium-doped periodically poled lithium niobate crystal; LPF: long-wavelength-pass filter; BS: pellicle beam splitter; PD: photo-detector.
(b) Optical spectra of the idler (grey), signal (blue) combs, measured with a Fourier transform spectrometer (Vertex 70, Bruker) at a resolution of 6 GHz. Here $f_{signal}$= 192.2 THz (1560 nm) and $f_{idler}$=89.64 THz (3344 nm). The resolution of the spectrometer is not high enough to resolve the individual comb lines, which line spacing is 300 MHz.

Each near-infrared signal comb is converted to the mid-infrared by difference frequency generation with a pump continuous-wave ytterbium-doped fiber laser system (YLR-15k-1064-LP-SF, IPG Photonics) of frequency $f_{pump}$ = 281.81 THz (1063.8 nm) in a temperature-controlled 40-mm-long MgO-doped periodically-poled lithium niobate (PPLN) crystal (MOPO1-0.5-40, Covesion). Both signal and pump beams are linearly s-polarized. At the crystals, the pulse duration of the signal combs is of the order of 300 ps. After the crystals, optical long-wavelength-pass filters (Edmund Optics; cutoff at 125 THz - 2.4 µm) filter out the light of the pump and of the signal. Two mid-infrared combs are generated (Fig. 1b): their center frequency of $f_{idler}$ and $f_{idler}$ +$f_{shift}$, respectively, is tunable between 85 and 95 THz (between 3150 and 3500 nm), where $f_{idler}$= $f_{pump}$- $f_{signal}$. Their line spacing of $f$ and $nf+\Delta f$, respectively, is the same as that of the near-infrared combs, chosen between 100 and 500 MHz. It only depends on electronic settings. The combs are remarkably flat-top. When the line spacing $f$ is 300 MHz, the idler comb spectrum spans 0.36

THz (13.4 nm) and it includes 1200 lines within 10-dB power variation. The average power of each idler comb exceeds 500 µW, when the power of the pump is 3 W and that of the signal 150 mW. The conversion efficiency of the PPLN crystals is about 0.28 mW/(W$^2$ cm), consistent with the specifications provided by the manufacturer. The power per idler comb line is about 420 nW.

A dual-comb interferometer is then realized. One mid-infrared comb is transmitted through a 70-cm-long gas cell with CaF$_2$ windows at Brewster angle. It is combined with the second mid-infrared comb, which acts as a local oscillator, on a 50:50 pellicle beam splitter. To avoid detector nonlinearities, the beam is attenuated to less than 100 µW per comb, before it is focused onto a fast photodetector (PVI-4TE-5, Vigo System SA) of 50-MHz bandwidth. The time-domain interference between the two combs, the interferogram, is digitized by an acquisition card (ATS9462, AlazarTech) at a rate of 180 10$^6$ samples s$^{-1}$. During the measurement of an interferogram, the frequency $f_{signal}$ of the near-infrared continuous-wave laser is kept unchanged. From one interferogram to another, it can be quickly tuned thanks to the frequency agility of the extended-cavity laser diode. The acquisition and computation of the interferograms are performed without phase corrections. No active stabilization is used either, which significantly reduces the experimental complexity. The Fourier transform of the interferogram reveals a radio-frequency comb, of a center frequency of $f_{shift}$ and a line spacing of $\Delta f$, with the imprint of the mid-infrared molecular absorption lines.

The conversion of the radio-frequency scale to the optical frequency scale in the spectra is performed *a posteriori.* For calibration of the optical frequency scale, the measurement of the optical frequencies $f_{pump}$ and $f_{signal}$ of the pump and signal near-infrared continuous-wave lasers, respectively is required, in addition to that of the line spacings $f$ and $f+\Delta f$. For coarse adjustment, we monitor such optical frequencies with a wavelength meter (WA-1000, Burleigh) of an accuracy of 100 MHz. For precise calibration, the frequencies of the two continuous-wave lasers are referenced to an active hydrogen maser (Kvarz CH1-75A) via a metrology erbium-doped fiber frequency comb. The two continuous-wave lasers are free running. For both, on a time-scale of one second, no significant drift is observed. The standard deviation of the frequency instabilities of each, measured against the metrology frequency comb, is smaller than 0.5 MHz at one second.

The spectroscopy instrument described above has been conceived to overcome the main difficulties of mid-infrared dual-comb spectroscopy. First, dual-comb spectroscopy requires that the coherence between the two combs be maintained during the time of the measurement of the interferogram. Otherwise artifacts severely distort the spectrum. In the near-infrared region, a variety of sophisticated solutions have been successfully employed with mode-locked lasers. They include stabilizing the combs against Hz-line-width continuous-wave lasers [7], or correcting for the relative drifts of the two lasers, either with analog electronics [24] or with digital processing [25]. Their implementation in the mid-infrared region is more involved.

Here we avoid such complexity by excellent passive relative coherence and common-noise rejection. The two near-infrared combs [23] are generated from a single continuous-wave laser and they are broadened in the same nonlinear fiber where they counter-propagate. The pump laser in the difference frequency generation process is common to the two combs. Second, the two mid-infrared combs have a flat intensity distribution. This overcomes the difficulty of the poor dynamic range of mid-infrared photo-detectors and of fast digitizers. Here, each radio-frequency comb line is measured at high signal-to-noise ratio, even for short recording times. Third, the moderate spectral span is also favorable to high sensitivity at short measurement times, as well as high refresh rates $\Delta f$ of the interferograms: in dual-comb spectroscopy, the signal-to-noise ratio and the refresh rate scale with the inverse of the number of spectral elements. For signal-to-noise ratio improvement by reduction of the spectral span in the dual-comb spectra, the output of mode-locked lasers is often spectrally filtered [7, 21]. Our technique, which inherently combines moderate spans as well as fast and mode-hop-free tunability of the spectral position, is a simple alternative to spectral filtering.

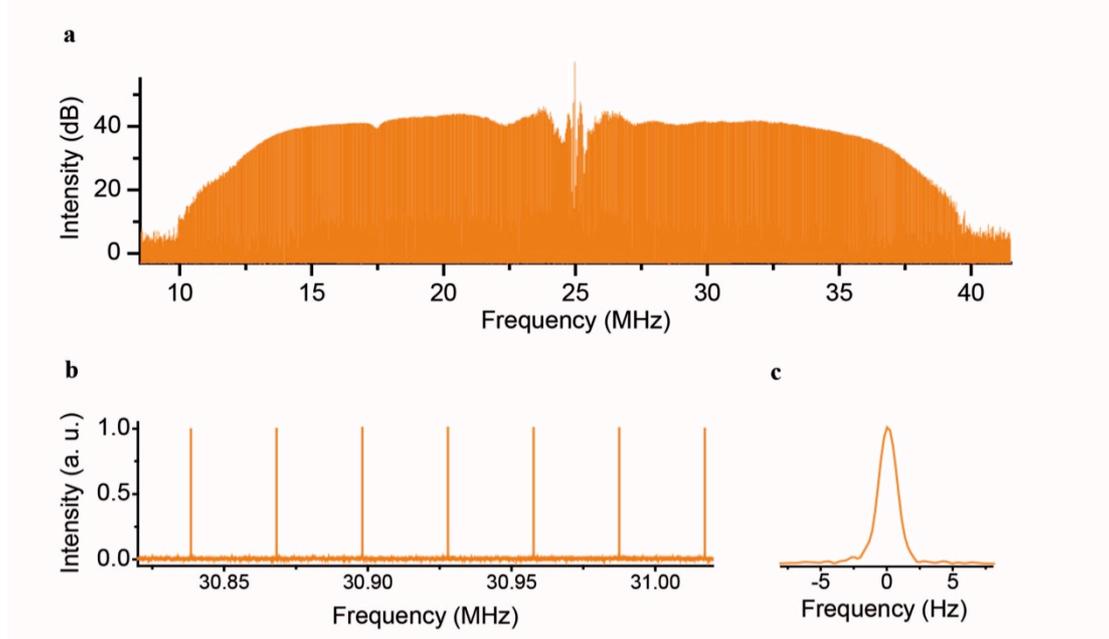

**Figure 2. Experimental dual-comb spectrum with resolved comb lines.**
(a) Radio-frequency spectrum of the detector signal on a logarithmic y-scale. The optical comb line spacing is $f$=230 MHz and the difference in line spacings is $\Delta f$=0.03 MHz. The spectrum results from a Fourier transform of 198 10$^6$ samples.
(b) Zoom on a few combs lines on a linear y-scale.
(c) In the radio-frequency spectrum, the comb line width is 0.9 Hz, limited by the recording time of 1.1 s.

We illustrate the excellent passive mutual coherence and the good signal-to-noise ratio in flat-top dual-comb spectra by performing recordings up to a time-scale of one second without any servo-control electronics, adaptive sampling or phase-correction. Figure 2 shows the radio-frequency dual-comb spectrum of the mid-infrared idlers, measured within 1.1 s. The center optical frequency is 89.65 THz, which is mapped in the radio-frequency range to $f_{shift}$=25 MHz. The line spacing in the optical domain is $f$ = 230 MHz and in the radio-frequency domain, it is $\Delta f$ = 0.03 MHz. More than 1050 comb lines, spanning 0.25 THz, are resolved with a transform-limited line-width of 0.9 Hz and a signal-to-noise ratio (SNR) exceeding 780. The SNR is calculated as the maximum of a comb line divided by the standard deviation of noise floor between two individual comb lines. Our data acquisition system is limited to a maximum sampling duration of 1.1 s. Within this time limit, the SNR scales with the square root of the measurement time. In Figure 2, the signal at the frequency $f_{shift}$=25 MHz is the residual of the carrier continuous-wave laser. Around $f_{shift}$, the spectral envelope shows some modulations. Such intensity fluctuations, which are due to the cross-phase modulation of the continuous-wave carrier with the pulses, are already present in the spectrum of each near-infrared signal comb, as shown in [23]. The use of electro-optic modulators with a higher extinction ratio would diminish such effect. Around 17.49 MHz (89.61 THz in the optical domain), one can distinguish the nitrogen-broadened $P$(3) blended manifold of the $\nu_3$ band of $^{12}CH_4$. The pressure of $CH_4$ is 13.5 Pa and the pressure of $N_2$ is 96 10$^3$ Pa.

Another significant advantage of our frequency comb generators without resonant elements is their freely adjustable line spacing. Their spacing may be selected to match the desired spectral resolution while the measurement times and the refresh rates of the interferograms are kept optimal. In the 3-μm region, for highly crowded spectral features at low pressure like $Q$ branches composed of Doppler-broadened lines, a line spacing $f$ of 100 MHz is desirable. We illustrate our high resolution of 100 MHz with a spectrum in the region of the $Q$ branch of the $\nu_{11}$ band of ethylene ($^{12}C_2H_4$). The difference in repetition frequencies $\Delta f$ is set to 15 kHz. The entire spectral span is 230 GHz and includes 2300 resolved comb lines. The recording time of an individual spectrum is 720 µs at a comb optical line width of 9.25 MHz. The spectrum of Fig. 3 results from 100 averaged individual spectra for a total measurement time of 72 ms. In Figure 3, a spectral portion of 75 GHz is displayed, where the maxima of the resolved comb lines are plotted. The 70-cm long single-pass cell is filled to a pressure of 140 Pa of ethylene in natural abundance at a temperature of 296 K. The Doppler-broadened full-width at half-maximum of the ro-vibrational lines is 215 MHz, but in the $Q$ branch, the high density of transitions generates only partly-resolved blended lines.

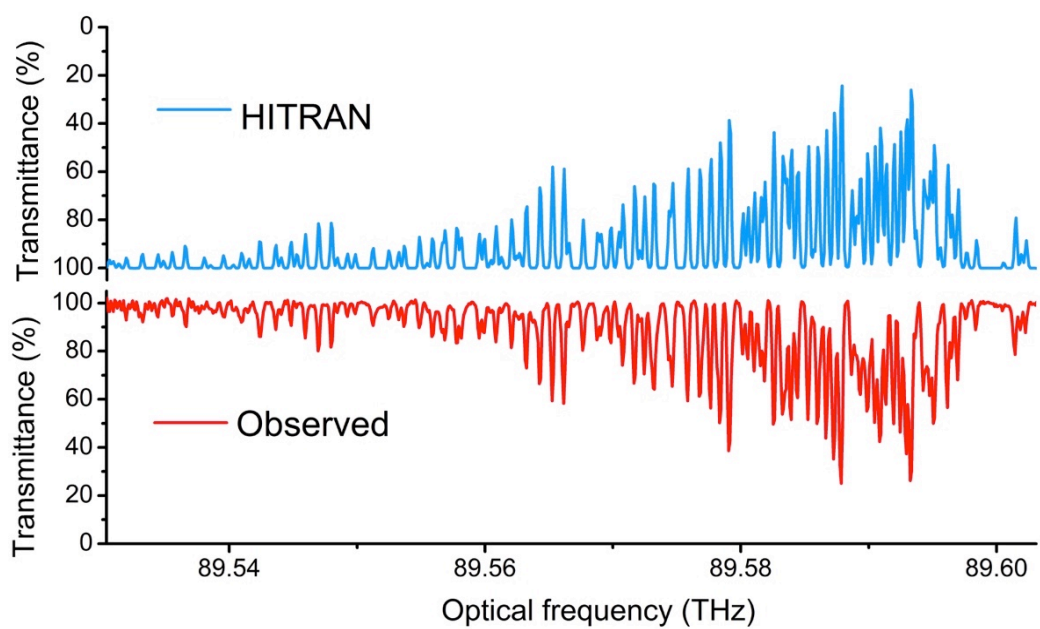

**Figure 3. Portion of a dual-comb transmission spectrum showing the crowded $Q$ branch of the $\nu_{11}$ band of ethylene.**
The transmittance signal is plotted at the comb line positions only. The comb line spacing is 100 MHz and the recording time is 72 ms. A good agreement between the experimental spectrum (red) and a spectrum (mirrored representation, blue) computed from the line parameters available in the HITRAN database [26] is obtained.

Spans broader than 0.3 THz may be achieved by tuning the frequency $f_{signal}$ of the signal continuous-wave laser in a step-wise way and stitching the resulting dual-comb spectra. Figure 4 illustrates Doppler-limited spectra of methane in the region of the $Q$ branch of the $\nu_3$ band. The comb line spacing $f$ is 115 MHz and the difference in repetition frequencies is $\Delta f$ =20 kHz. Methane in natural abundance is contained in the 70-cm-long gas cell with a pressure of 11.4 Pa at a temperature of 296 K. In Fig. 4a, three spectra, each measured within 72 ms at a resolution of 115 MHz partly overlap. In each, the signal-to-noise ratio SNR is 350 and more than M=1800 comb lines are resolved. Therefore the signal-to-noise ratio per unit of time exceeds 1300 $s^{-1/2}$ and its product with the number of comb lines is 2.3 $10^6 s^{-1/2}$. Such value is higher than that reported in [21] where combs stabilized against hertz line-width lasers were used. Fig. 4 b displays a portion of the stitched spectrum, spanning 185 GHz, with well-resolved Doppler-broadened methane lines and a normalized baseline.

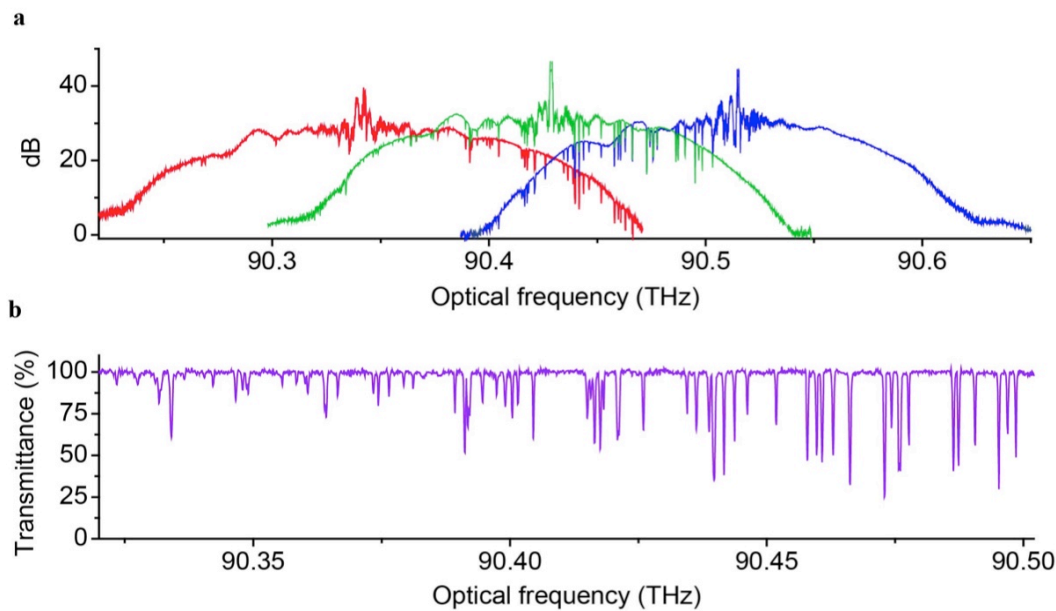

**Figure 4. Dual-comb spectra of the *Q* branch of the $\nu_3$ band of $^{12}CH_4$ at a resolution of 115 MHz.**
(a) Three experimental spectra (red, green, blue) with a central frequency of 90.34 THz, 90.42 THz and 90.51 THz, respectively.
(b) Portion of the transmittance spectrum showing the manifolds *Q*(2) to *Q*(12).

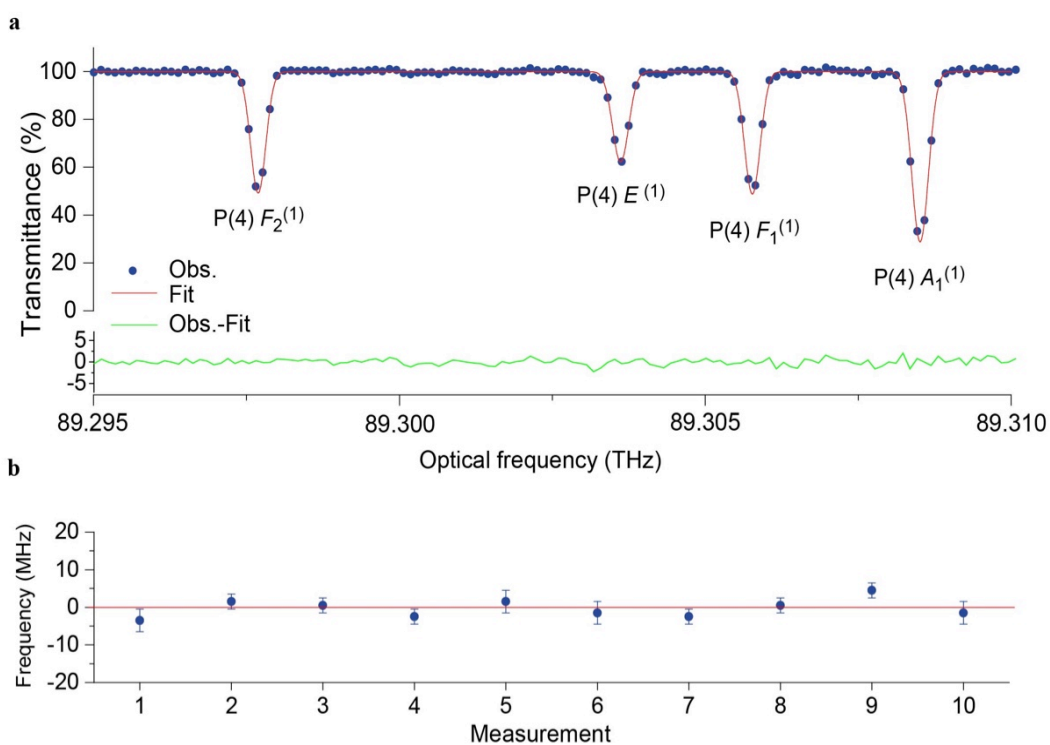

**Figure 5. Line parameter measurements.**
(a) Portion of a dual-comb spectrum showing the *P*(4) manifold of the $\nu_3$ band of $^{12}CH_4$. The pressure of $CH_4$ in natural abundance is 11.4 Pa and the absorbing path is 70 cm. Gaussian profiles (red) fit the experimental lines (blue dots). The residuals "Obs.-Fit" (green) do not show any systematic deviation and their standard deviation is 0.7%.
(b) Statistical distribution of 10 measurements of the $P(4)F_1^{(1)}$ line center frequency. The error bar for each measured frequency is the standard error of the fitted line position parameter. The reduced chi-square of the fit is 1.22. The horizontal red line represents the average center frequency (89305771.5 MHz), which has been subtracted from each measured frequency for clarity.

As any technique of Fourier transform spectroscopy [27], dual-comb spectroscopy is prone to subtle artifacts, which can distort the molecular line shapes and provide inaccurate molecular parameters. For assessing our experimental line parameters, we fit (Fig.5a) a Doppler profile to the tetrahedral sub-level lines $F_2^{(1)}$, $E^{(1)}$, $F_1^{(1)}$, $A_1^{(1)}$ in the $P(4)$ manifold of the $\nu_3$ band of $^{12}CH_4$. The residuals between the experimental spectrum and its fit have a standard deviation of 0.7%. Importantly, the instrumental line-shape is negligible, because narrow comb lines sample the molecular spectrum. The calibration of the frequency scale is performed against the hydrogen maser. The statistical distribution of the 10 frequency measurements for the $P(4)F_1^{(1)}$ line position is shown in Fig.5b as an example. The statistical uncertainty amounts for 0.8 MHz. The frequencies $f_{pump}$ and $f_{signal}$ of the two continuous wave-lasers are each given a conservative uncertainty of 0.5 MHz, which corresponds to their measured instabilities at one second. The drifts of the two continuous-wave lasers represent the main source of non-statistical uncertainty. If better accuracies were sought for, the two lasers could straightforwardly be actively stabilized. The self-induced pressure shifts for the $P(4)$ manifold have not been reported in the literature to our knowledge. On the basis of measurements in the $Q$ branch [28], we infer that they are, at 11.4 Pa, significantly smaller than -100kHz and we include them in our uncertainty. We neglect the light shift, which has been reported to (–0.0039 ± 0.0049) kHz/μW for the $P(7)F_2^{(2)}$ line [29]. We average the ten frequency measurements shown in Fig.5b and measure the center frequency of the $P(4)F_1^{(1)}$ line to 89305771.5(8) MHz. The number within parenthesis is the 1σ uncertainty, including statistical and systematic effects, in units of the last digits. Interestingly, we reach an accuracy on the MHz scale for spectra measured within less than one second. We compare our measurements with Doppler-free measurements based on saturated absorption spectroscopy [29] (accuracy of around +/- 2 kHz) and Doppler-limited measurements based on accurate dual-comb spectroscopy [21] (accuracy of +/- 300 kHz). The agreement, within 100 kHz, is significantly better than our accuracy. For line intensity measurements, we estimate our overall uncertainty to 4%, mostly due to the uncertainty on the absolute calibration of the employed capacitance manometer. The line intensities agree within 2.4 % with those measured by [30] of an accuracy of 2%. More details on the measured positions and intensities of the lines in the $P(4)$ manifold are given in Table 1 and Table 2, respectively.

| Assignment | Frequency (MHz) | | | | |
|---|---|---|---|---|---|
| | This work | [29] | This work - [29] | [21] | This work - [21] |
| $P(4)\ F_2^{(1)}$ | 89 297 694.9(9) | 89 297 695.263 5  (20) | -0.4 | 89 297 695.06 (30) | -0.2 |
| $P(4)\ E^{(1)}$ | 89 303 617.9(11) | 89 303 620.049 2 (20) | -2.1 | 89 303 619.87 (30) | -2.0 |
| $P(4)\ F_1^{(1)}$ | 89 305 771.5 (8) | 89 305 771.609 3 (20) | -0.1 | 89 305 771.40 (30) | 0.1 |
| $P(4)\ A_1^{(1)}$ | 89 308 513.3(8) | 89 308 512.256 8 (20) | 1.0 | 89 308 512.19 (30) | 1.1 |

**Table 1.** Center frequencies of the lines in the $P(4)$ manifold of the $\nu_3$ band of $^{12}CH_4$ measured in this work and comparison with the accurate measurements reported in [29] by Doppler-free saturated absorption spectroscopy and [21] by Doppler-limited dual-comb spectroscopy.

| Assignment | Line intensity (cm.molecule$^{-1}$) | | |
|---|---|---|---|
|  | This work | [30] | (This work –[30])/ [30] (%) |
| $P(4)\ F_2^{(1)}$ | $7.13(13)\times10^{-20}$ | $7.310(147)\times10^{-20}$ | -2.41 |
| $P(4)\ E^{(1)}$ | $4.80(13)\times10^{-20}$ | $4.887(98)\times10^{-20}$ | -1.86 |
| $P(4)\ F_1^{(1)}$ | $7.24(13)\times10^{-20}$ | $7.339(147)\times10^{-20}$ | -1.28 |
| $P(4)\ A_1^{(1)}$ | $1.206\ (20)\times10^{-19}$ | $1.227(25)\times10^{-19}$ | -1.71 |

**Table 2.** Individual intensities of the lines of the *P*(4) manifold of the $\nu_3$ band of $^{12}CH_4$ measured in this work. Comparison with the data published in [30]. The temperature is 296K and the intensities are given corrected for isotopic abundance (0.98893 for $^{12}CH_4$ [30]).

    Fifteen years ago, combs based on electro-optic modulators had been harnessed [31] for a proof-of-principle demonstration of dual-comb optical coherence tomography. In the past two years, dual-comb spectroscopy with electro-optic modulators has become a promising trend [23, 32-36] for multiplex spectroscopy. So far all the demonstrations have been performed [23, 32-36] in the telecommunication near-infrared region. Here, we have extended the operation of dual-comb spectroscopy with electro-optic modulators to the mid-infrared range, with first results in the 3-µm region of the CH, OH and NH stretches in molecules. Our technique shows several distinguishing advantages compared to that which use systems based on mode-locked lasers. Line spacing and spectral position can be selected quickly and freely by simply dialing a knob. Moreover, although systems with a single electro-optic modulator [32-34, 36] only generate a limited number of comb lines, our combination of an intensity modulator and a non-linear fiber easily produce lines in excess of 1200, well suited for simultaneously interrogating several tens of transitions at Doppler-limited resolution within less than one hundred milliseconds. An optimized design of the nonlinear fiber used for spectral broadening of the near-infrared combs should increase the number of comb lines to an excess of 5000. The flat spectral envelope of our mid-infrared combs favors a high signal-to-noise ratio for all comb lines, even at very short measurement times. Our system is designed to generate two combs which are passively mutually coherent. We have shown that such design makes it possible to resolve comb lines of a Fourier-limited width narrower than one hertz without active stabilization or phase-lock electronics. Active stabilization may be required for long averaging times but it would then only require easy-to-implement low-bandwidth feedback loops. The spectrometer can easily be engineered to be compact and portable. The simplicity of our scheme is particularly meaningful in the mid-infrared region, where other laser systems based on nonlinear frequency conversion, like optical parametric oscillators, may be extremely challenging to suitably stabilize. Our technique of mid-infrared multiplex spectroscopy over moderate spectral spans offers unprecedented opportunities for rapid measurements of accurate molecular line parameters. The instrumental line-shape is negligible thanks to the resolved narrow comb lines. In our set-up, replacing the pump laser of fixed optical frequency by a tunable continuous-wave laser will extend the spectral range of our spectrometer further in the mid-infrared domain. For future applications to e.g. trace gas detection, high finesse resonators [9,36] can enhance the sensitivity to weak absorptions. Laboratory spectroscopy of transient species and, with further system development, in-field detection of e.g. pollutants and molecules of atmospheric relevance may benefit from our novel instrument.

**Acknowledgments**
This research has been funded by the European Research Council (Advanced Investigator Grant 267854), the Munich Center for Advanced Photonics, the Max Planck Foundation, IXCORE Fondation pour la Recherche, PARI PHOTCOM Région Bourgogne, Labex ACTION program (Contract No. ANR-11-LABX-0001-01).